# Postprandial morphological response of the intestinal epithelium of the Burmese python (*Python molurus*)

Jean-Hervé Lignot[a,*], Cécile Helmstetter[a] and Stephen M. Secor[b,c]

[a]CNRS, Centre d'Ecologie et Physiologie Energétiques,
23 rue Becquerel, F-67087 STRASBOURG cedex 2, France.

[b]Department of Physiology
University of California, Los Angeles
School of Medicine
Los Angeles, CA 90095-1751
USA

[c]Present address: Department of Biological Sciences
Box 870344
The University of Alabama
Tuscaloosa, AL 35487-0344
USA

* Corresponding author.
CNRS, Centre d'Ecologie et Physiologie Energétiques,
23 rue Becquerel, F-67087 STRASBOURG cedex 2, France.
E-mail address: J-H.Lignot@c-strasbourg.fr (J-H. Lignot)

**Abstract**

The postprandial morphological changes of the intestinal epithelium of Burmese pythons were examined using fasting pythons and at eight time points after feeding. In fasting pythons, tightly packed enterocytes possess very short microvilli and are arranged in a pseudostratified fashion. Enterocyte width increases by 23% within 24 h postfeeding, inducing significant increases in villus length and intestinal mass. By 6 days postfeeding, enterocyte volume had peaked, following as much as an 80% increase. Contributing to enterocyte hypertrophy is the cellular accumulation of lipid droplets at the tips and edges of the villi of the proximal and middle small intestine, but were absent in the distal small intestine. At 3 days postfeeding, conventional and environmental scanning electron microscopy revealed cracks and lipid extrusion along the narrow edges of the villi and at the villus tips. Transmission electron microscopy demonstrated the rapid postprandial lengthening of enterocyte microvilli, increasing 4.8-fold in length within 24 h, and the maintaining of that length through digestion. Beginning at 24 h postfeeding, spherical particles were found embedded apically within enterocytes of the proximal and middle small intestine. These particles possessed an annular-like construction and were stained with the calcium-stain Alizarine red S suggesting that they were bone in origin. Following the completion of digestion, many of the postprandial responses were reversed, as observed by the atrophy of enterocytes, the shortening of villi, and the retraction of the microvilli. Further exploration of the python intestine will reveal the underlying mechanisms of these trophic responses and the origin and fate of the engulfed particles.

**Introduction**

The small intestine possesses the adaptive capacity to alter form and function in response to changes in digestive demand (Piersma and Lindström, 1997). Changes in diet composition, meal size, and meal frequency have all been demonstrated to induce alterations of the morphology and/or function of the intestinal epithelium. Examples of intestinal plasticity among mammals include the doubling of intestinal D-glucose uptake by mice switched from a low carbohydrate to a high carbohydrate diet, the 50% increase in intestinal mass of mice that have increased food intake during cold exposure and/or lactation, and the atrophy of the small intestine of ground squirrels during the long-term fast of hibernation and subsequent intestinal hypertrophy with feeding following spring emergence (Carey, 1990; Diamond and Karasov, 1984; Hammond et al., 1994).

Although important for improving nutrient uptake and energy flux, these mammalian responses are nevertheless modest in scope when compared to the phenotypic responses displayed by the intestines of snakes that feed infrequently in the wild (Secor and Diamond, 2000). Snakes that employ the sit-and-wait tactic of prey capture (pythons, boas, vipers, and rattlesnakes) feed at intervals of one to several months and consume prey that may even exceed their own body mass (Greene 1992; Pope, 1961; Secor and Diamond, 2000). With each meal, these snakes double the mass of their small intestine and increase rates of intestinal nutrient transport by 5 to 20-fold (Secor and Diamond, 2000). Upon the completion of digestion, intestinal mass is reduced and nutrient transport is downregulated. Proposed to benefit these snakes energetically, this capacity to broadly regulate intestinal morphology and performance must involve steps that remodel epithelial structure and function (Secor, 2005).

Much of the attention on the digestive physiology of snakes has recently been directed at the sit-and-wait foraging Burmese python, *Python molurus*. Studies have demonstrated for this snake the rapid upregulation of intestinal function after feeding, a graded functional response with increased meal size, a compensatory increase in intestinal function following resection of the small intestine, and the dependency upon luminal proteins to trigger intestinal upregulation (Secor and Diamond, 1995; 1997; Secor et al., 2000; 2002). These accounts, as well as the study of  Starck and Beese (2001) have briefly described the morphological changes that the python's intestinal epithelium faces with fasting and feeding.

Given the dependency of epithelial structure on intestinal function for the Burmese python, it is necessary to examine in more detail the postprandial changes of the python intestinal epithelium. Our long-term goals are to identify for the python the cellular mechanisms responsible for modulating intestinal morphology and function. The objectives of this study are: (1) to track the postfeeding changes in intestinal mass, enterocyte size, and microvillus dimensions; (2) to view structural changes of the villus surface during digestion; (3) to identify the effect of intestinal position on epithelial morphology; and, (4) to describe unique phenomena of the python enterocytes with feeding or fasting. Highlights of this study include the postprandial hypertrophy of the villus surface, the rapid lengthening of the intestinal microvilli, and the presence of particles (possibly bone fragments) embedded in the brush border membrane.

**Materials and methods**

*Animals*

Burmese pythons are native to the subtropical regions of southeast Asia and feed largely on birds and mammals (Pope, 1961). We purchased hatchling pythons (~ 100 g) from commercial breeders (Captive Bred Reptiles, Oklahoma City, OK and Zooland, Strasbourg, France) and maintained each snake in an individual cage within a temperature-controlled room (28-30°C) under a 14h:10h light-dark cycle. Pythons were fed laboratory rats bi-weekly with each meal mass being approximately 25% of the snake's body mass. Prior to experiments, snakes were fasted for 30 days to ensure that they were postabsorptive (Secor and Diamond, 1995). During this fasting period, snakes continued to have access to water. Snakes were killed by severing their spinal cord immediately posterior to the skull following the 30-day fast or at 0.25, 0.5, 1, 2, 3, 4, 6, and 14 days following their consumption of rat meals equalling 25% of their body mass. For each sampling period, intestinal samples were collected from a minimum of three snakes. The 49 snakes used in this study averaged (mean ± 1 SE) 775 ± 44 g and ranged in age between 0.5 and 2 years old. All housing and experimental procedures were conducted under approved institutional animal care and use protocols at the University of California, Los Angeles, the University of Alabama, and the Centre National de la Recherche Scientifique.

*Light microscopy*

Following the severing of the spinal cord, mid-ventral incisions were made along the length of the snake's body. The small intestine was removed, flushed of any contents with ice-cold reptilian ringers solution, weighed, and divided into equal-lengths thirds (proximal, middle, and distal). Each third was weighed and segments were removed from the middle of each third for histological examination. One-cm segments, fixed in either 10% neutral-buffered formalin solution or 3% paraformaldehyde in buffered saline, were dehydrated and embedded in paraffin. Intestinal cross-sections (6 µm) were stained with hematoxylin and eosin for morphometric measurements or stained with Alizarine red S to highlight calcified tissue. Using samples from three snakes for each of seven sampling periods [fasted, 0.25, 0.5, 1, 3, 6, and 14 days postfeeding (DPF)], we measured the width of the mucosa and muscularis/serosa layers of each intestinal segment using a light microscope and video camera linked to a computer and image-analysis software (Motic Image Plus, British Columbia, Canada). From these same segments, we also measured the height and width of ten enterocytes, from which we calculated enterocyte volume (based on the formula for a cylinder). For each snake, enterocyte height, width, and volume were averaged for each intestinal segment.

*Transmission electron microscopy*

Small samples of intestinal mucosa were fixed in 2.5% glutarhaldehyde in 0.05M cacodylate buffer (pH = 7.4) for transmission electron microscopy. Samples were subsequently post-fixed in 1% osmium tetroxide, dehydrated in a graded series of ethanol, and embedded in either Spurr or Araldite 502 resin. Semi-thin ($\approx$ 1.5 µm) and ultra-thin sections ($\approx$ 90 nm) were respectively placed on poly-L-lysine-coated slides or copper grids. Semi-thin sections were stained with toluidin blue and observed to describe postprandial changes in enterocyte structure and organization. Ultra-thin sections were stained with uranyl acetate for 30 min and lead citrate for 3 min and examined on a Jeol JEM-100 CX electron microscope. For each intestinal sample, we selected four to five areas of microvillus to photograph at a magnification of X7,200. From these prints, the length and width of 5 – 10 microvilli were measured, selecting only microvilli cut along the central plane of their long axis. Actual microvillus length and width was calculated by dividing measurements by the total print magnification (X21,600). From prints taken of cross-sectioned microvilli, the density of microvilli was calculated by counting the number of microvilli within a 1μm² area. We measured microvillus dimension and density of the proximal small intestine from 3 - 6 snakes per sampling period, and also from the middle and distal segments of three snakes fasted, at 1 and 6 days following feeding. For each snake and intestinal position, we used average microvillus length, width and density to calculate overall surface area of the microvillus membrane per μm² of flat epithelial surface using the formula for the surface area of a cylinder.

*Scanning electron microscopy*

We observed samples of the proximal intestine of fasted and fed pythons using conventional SEM. Samples were fixed for 2 h at room temperature in 3% paraformaldehyde in 0.05M phosphate buffer (pH = 7.4) and subsequently dehydrated through a graded series of ethanol. Samples were then rapidly bathed in 1,1,1,3,3,3-hexamethyldisilazane, air-dried, and attached to specimen stubs with carbon adhesive tabs or silver paint. Samples were gold coated (Edwards Sputter Coater) and examined with a Philips XL-30 ESEM using the conventional mode (high vacuum) and a Thornley-Everhart secondary detector. To observe samples in the environmental mode of the XL-30 ESEM, proximal intestinal samples were fixed in 3% paraformaldehyde and buffered saline for 2h at room temperature, washed, and immediately viewed. We held the relatively humidity at the sample surface at 80% by maintaining the sample chamber temperature and pressure at 4°C and 5 Torr, respectively. Chamber pressure was regularly decreased to 1 Torr in order to reduce the relative humidity to below 15%.

*Statistical analyses*

We used analysis of covariance (ANCOVA), with body mass as a covariate, to test for the effects of sampling time on intestinal mass, mucosa and serosa width, enterocyte length, width, and volume, and microvillus length, width, density, and surface area. A repeated-design analysis of variance (ANOVA) was used to test for positional effects (proximal, middle, and distal regions of the small intestine) on segment mass and enterocyte and microvillus dimensions. In conjunction with ANCOVA and ANOVA, we made post hoc planned pairwise mean comparisons between sampling periods. Our data are presented as means ± 1 SE, and we set the level of statistical significance at $P < 0.05$. Statistical analyses were conducted using the software package Statistica (StatSoft, Inc., Tulsa, Oklahoma).

**Results**

*Small intestinal mass*

The small intestine of fasted Burmese python represented one-third in mass of their gastrointestinal tract (esophagus, stomach, small intestine, and large intestine), and 2.5% of total body mass. With feeding, the mass and morphology of the small intestine changed dramatically. Within 12 h after feeding, total small intestinal mass has increased significantly ($P = 0.019$) by 47% (Fig. 1).The small intestine continued to grow thereafter, peaking at day 3 of digestion at almost twice its original fasting mass. For fasted snakes and at 14 DPF, we found no significant differences in mass among the proximal, middle, or distal intestinal segments. For the other sampling times (0.25 to 6 DPF), the proximal and middle segments were typically heavier than the distal segment (Fig. 1).

Fig. 1

*Small intestinal epithelium*

Conventional SEM revealed that intestinal villi of fasted pythons appeared punctuated due the sunken position of many cells (Figs. 2a, 2b). At this time, cell edges especially at the tip of the villi were well defined and there was no evidence of any cell being shed (Fig. 2c). One day after feeding, the surface of the villi had changed due to the hypertrophy of cells and the lengthening of the microvilli (Figs. 2d, 2e). Beginning at day 3 of digestion, longitudinal cracks formed along the narrow edges of the villi and at the villus tips, allowing cell contents to be released into the lumen (Figs. 2f, 2g, 2h). By day 6, the longitudinal cracks were less evident and the surface had become smoother in appearance (Fig. 2i). By day 14, the villi possessed a pitted appearance similar to that found in snakes fasted for 30 days (Fig. 2j).

Fig. 2

Environmental SEM, which allowed us to observe hydrated samples, demonstrated the short stature of the villi of fasted snakes and the presence of mucus bridges between villi (Figs. 3a, 3b, 3c). One day after feeding, we observed droplets attached to the tips of villi of the proximate and middle intestine (Figs. 3d, 3e), although such droplets were absent in the distal region (Fig. 3f). Three days after feeding, droplets increased in size and number, but were still only confined to the proximal and middle intestine (Figs. 3g, 3h, 3i). During continuous monitoring, several droplets on the villi were observed to fuse (Fig. 3j, 3k). We suspect that the droplets are composed of lipids, rather than being aqueous, because when samples were allowed to dehydrate, the droplets remained turgid (Fig. 3l).

Fig. 3

Fig. 4

Fig. 5

From light microscopy samples, we found intestinal enterocytes of fasted pythons to be tightly packed together, occasionally overlapping in a pseudostratified manner (Fig. 4a). With feeding, enterocytes increased in girth, resulting in a single, non-overlapping, layer of cells (Fig. 4b). Within 24 h after feeding, enterocytes of the proximate and middle small intestine had increased (P's < 0.034) in width by 23%, while at the same time decreasing (P's < 0.011) in length by 20% (Figs. 5a, 5b). By day 6 of digestion, proximal and middle enterocytes had expanded in width by 40%, their lengths have returned to fasting values, and their volumes had increase (P's < 0.013) by 80% (Figs. 5a, 5b, 5c). By day 14, enterocyte length, widths, and volumes had returned to fasting values. For several sampling periods, enterocyte dimensions varied significantly with intestinal position with a trend of a decrease in length, width, and thus volume distally (Figs. 5a, 5b, 5c). The postprandial widening of enterocytes resulted in a corresponding lengthening of the villi, as observed by the respective 38% and 47% increase (P's < 0.002) in the width of the mucosal layer of the proximate and middle intestinal regions (Fig. 6). In contrast, the width of the muscularis/serosa layer did not change for any of the intestinal segments with feeding (Fig. 6).

Contributing to the postfeeding increase in enterocyte size is their accumulation of lipid droplets. We first observed lipid droplets within cells of the proximate and middle intestine 24 h after feeding (Fig 7a). Droplets were spherical in shape and averaged $1.56 \pm 0.05$ µm (n = 500) in diameter, with a maximum diameter of approximately 5 µm. Lipid droplets were more heavily concentrated in cells at the villus tips, and occasionally occurred in cells along the full length of the villi. By day 4 of digestion, the occurrence of droplets had begun to decrease such that by day 6, small lipid droplets were found in only a few specimens, usually along the basal membrane of the enterocytes (Fig. 7b). We did not find lipid droplets in samples collected at 14 DPF, nor were any observed in samples taken from the distal intestine at anytime. Rather, the distal epithelium contained a large population of goblet cells (Fig. 7c).

*Small intestine brush border*

One of the most dramatic postprandial responses of the python's small intestine is the lengthening of their microvilli (Fig. 8a-g). For the fasted python, microvilli are relatively stunted, averaging for the proximate region $0.49 \pm 0.04$ µm in length (Fig. 9a). Within 6 h after feeding, the proximate microvilli had already doubled (P = 0.0008) in length. Thereafter, the microvilli continued to lengthen to peak at 24 h postfeeding at $2.37 \pm 0.18$ µm, 4.8-fold of fasting values. Intestinal microvilli remained lengthened through day 6, and had declined to fasting lengths by day 14. While increasing in length, the microvilli decreased in width, reducing their diameter by 14% within 12 h after feeding (Fig. 9b). The microvilli remained thinner for the remainder of digestion (until day 6), before returning to fasting widths by day 14. We detected no significant variation among sampling times in microvillus density for the proximate small intestine; density averaged $38.2 \pm 1.0$ microvilli per $\mu m^2$ throughout the study (Fig. 9c). Surface area of the proximate microvilli varied significantly ($P < 0.0001$) among sampling periods, almost doubling within 6 h after feeding (Fig. 9d). By 24 h postfeeding, one $\mu m^2$ of flat luminal surface was occupied by approximately 30 $\mu m^2$ of microvillus surface, 3.8 times the fasting condition. Microvillus surface area remained elevated through digestion and returned to fasting values by day 14. Only microvillus length varied significantly among intestinal positions. For fasted snakes, distal microvilli were longer (P's $< 0.02$) than the microvilli of the proximate and middle regions, whereas at 1 DPF the proximate microvilli were longer ($P = 0.0005$) than the distal microvilli.

An intriguing finding for many of the postfeeding samples was the presence of large spherical particles ($3.30 \pm 0.20$ $\mu m$ diameter, $n = 60$) partially embedded in the brush border membrane. These particles were first observed at day 1 of digestion and were found scattered along the proximate and middle intestine until day 6. They were easily observed with light microscopy, SEM, and TEM, the latter illustrating for many particles their annular construction and that they were surrounded by microvillus-bearing membrane (Figs. 10a, 10b, 10c). The positive staining of the particles with Alizarine red S indicates strongly that they are composed of calcium (Fig. 10c).

Fig. 10

**Discussion**

Burmese pythons modulate intestinal morphology with each meal, as highlighted by the postprandial doubling of mucosal mass and almost 5-fold lengthening of the microvilli, followed by the atrophy of enterocytes and retraction of the microvilli once digestion is completed. This set of responses by python mucosa (with the exception of the microvilli) is not unlike that experienced by other vertebrates. Fasting-induced atrophy of the intestinal epithelium has been observed for fishes (McLeese and Moon, 1989), amphibians (Cramp and Franklin, 2003), reptiles (Secor and Diamond, 2000), birds (Hume and Biebach, 1996), and mammals (Dunel-Erb et al., 2001). In these cases, fasting results in an overall decrease in mucosal mass, a shortening of the villi, and to some extent, a reduction of enterocyte width. Feeding rapidly reverses the fasting response; intestinal mass and luminal surface area increases and enterocytes swell. The magnitude by which the intestine modulates its mass with feeding and fasting varies among species. The postprandial 100% increase in intestinal mass for the python is similar in magnitude to that experienced by other infrequently-feeding snakes and by active and feeding thirteen-lined ground squirrels (*Spermophilus tridencemlineatus*) following their emergence from hibernation (Carey and Sills, 1992; Secor and Diamond, 2000). In contrast, feeding induced more modest increases in intestinal mass (25 – 50%) for frequently-feeding snakes, other reptiles, and amphibians (Secor and Diamond, 2000; Secor, 2005).

*Enterocyte hypertrophy*

The feeding and fasting-related changes in intestinal mass of the python is largely due to changes of the mucosal layer, as the muscularis/serosa width and mass are unaltered by feeding (this study; Secor and Diamond, 1995). For fasting pythons, enterocytes, the chief absorptive cells of the intestinal epithelium, are tightly packed together with cells overlapping each other in a pseudostratified arrangement. As also observed by Starck and Beese (2001), feeding results in a reshaping of python enterocytes in such a way that they cease to overlap and become arranged in a single layer of columnar cells that typifies the epithelium of the vertebrate intestine. Enterocytes also swell in size within 1 day after feeding, as revealed in this study by SEM. Enterocyte width and volume continued to increase during digestion, before peaking at day 6 when digestion was close to completion (Secor, 2003). During this time, the villi experienced several structural changes; longitudinal cracks appear along the villus edges by day 3 and then largely disappear by day 6. By day 14, almost a week after digestion has been completed, enterocytes had returned to their fasting appearance.

For mammals, luminal, secretory, hormonal, and neural signals have each been implicated to stimulate intestinal response, including mucosal growth (Gershon et al., 1994; Johnson and McCormack, 1994; Solomon, 1994; Walsh, 1994). For the Burmese python, it has been demonstrated that the sight and smell of food, the stretching of their stomach using a balloon catheter, the luminal presence of pancreaticobiliary secretions, and the intestinal infusion of saline and glucose solution each fail to generate a morphological response by the intestine (Secor, et al., 2002). In contrast, the infusion of solutions containing amino acids and peptides into the small intestine results in rapid mucosal hypertrophy (Secor et al., 2002). In spite of these findings, the mechanisms responsible for the postprandial hypertrophy of the python's enterocytes, and their subsequent atrophy once digestion has completed still remain unclear. Do the amino acids stimulate cellular growth, or together with absorbed monosaccharides and lipids, do they simply fill the cells thereby increasing their volume? A potential active mechanism that might be employed, but has not yet been demonstrated, is the mobilization of amino acids from body stores and their delivery via circulation to the intestinal epithelium. Within the enterocytes, the amino acids could be used to construct cellular components that both enlarge cells and enhance function.

An alternative mechanism of villus hypertrophy proposed by Starck and Beese (2001) involves the pumping of fluid into the villi from their capillary network, a mechanism supported by the postprandial 11-fold increase in mesenteric blood flow for the Burmese python (Secor and White, 2004). As suggested, the villi elongate due to the increase in hydrostatic pressure, and the compressed pseudostratified enterocytes lining the villi therefore expand in a fashion similar to that which occurs with the filling of the urinary bladder. For the Burmese python the fluid-pumping hypothesis is questionable considering that segments of python intestine surgically isolated from contact with luminal nutrients, but still receiving vascular supply, failed to exhibit villus hypertrophy after feeding, even though the remaining intact intestine which had received nutrients did experienced hypertrophy (Secor et al., 2000). Still, it might be a combination of mechanisms involving both luminal and vascular contributions that is responsible for enterocyte growth.

The fasting-related atrophy of the python's intestinal mucosa could simply occur as a reversal of the mechanisms involved in its growth. The cellular elements that enlarge enterocytes either are absorbed by the blood stream or expelled luminally. Alternatively, the fasted mucosa is composed of a fresh set of newly-proliferated enterocytes that arose following  the completion of digestion, and which remains dormant and tightly packed until the snake eats again (Starck and Beese, 2001).

*Enterocyte lipid accumulation*

Lipid droplets within enterocytes have been observed for recently-fed fish (Noaillac-Depeyre and Gas, 1974, 1979) and during the ontogenetic transition between the lecitotrophic and exotrophic larval stages of fishes (Sire et al., 1981; Diaz et al., 1997; Garcia Hernandez et al., 2001). For the python, we suspect that the source of their enterocyte lipids is their rodent meals, which are estimated to be about 15% fat. While the accumulation of lipid droplets by python enterocytes during digestion undoubtedly contributes to their hypertrophy, it has also been found that diets lacking lipids can also generate mucosal growth (Secor et al., 2002). In addition, there was considerable variation in the presence of lipids droplets, droplets were more concentrated in enterocytes positioned at the tips and along the edges of the villi. Finding in this study that lipid droplets only occurred in enterocytes of the proximal and middle intestine, suggests that all of the lipids of the rodent meals are absorbed by the anterior two-thirds of the small intestine. Once absorbed, the lipid droplets move through the enterocytes, and subsequently pass across the basal lateral membrane into circulation. Towards the end of meal digestion (6 DPF), lipid droplets were relatively smaller and concentrated at the basal end of the enterocytes.

*Modulation of microvillus length*

Microvilli of the fasted python are very much stunted, and with the 400% increase in length within 24 h after feeding, they attain the length normally observed for the microvilli of other species, both fasted and fed (Secor, 2005). Similar rapid lengthening of the intestinal microvilli (doubling within 24 h) has been observed for rats whose microvilli had been damaged by an acute exposure to kidney bean lectin (Weinman et al., 1989). To the best of our knowledge, the dramatic modulation of microvillus length simply to feeding has not been documented for any other animal species and thus may exist as an adaptive trait for Burmese pythons. Pythons, and other infrequently-feeding snakes, up and downregulate intestinal performance, respectively, with feeding and fasting (Secor and Diamond, 2000). It has been proposed that the fasting-related downregulation of gastrointestinal performance serves to reduce daily energy expenditure between the infrequent meals (Secor, 2001). Thus for the python, a mechanism by which intestinal performance is regulated is by changing surface area by lengthening and shortening the microvilli. As the microvilli lengthen after feeding, intestinal capacity to hydrolyze and transport nutrients concurrently increases, and as the microvilli shorten with fasting, intestinal performance is downregulated (Secor, 2005).

*Brush border particles*

Beginning at 24 h postfeeding, the brush border of many samples collected from the proximal and middle small intestine were embedded with spherical particles. Because many of these particles were surrounded by microvilli-bearing membrane, it appeared that they had been engulfed in a phagocytotic fashion. We are not aware of any other report of the intestinal epithelium actively importing luminal material of such large size (up to 6 μm in diameter). The pinocytotic absorption of intact macromolecules, including dietary proteins and immunoglobulins, has been described for infant humans and neonatal rats and pigs (Rodewald, 1970; Walker, 1986; Weaver and Walker, 1989). These macromolecules enter between microvilli and may remain intact as they transverse the enterocytes, thus providing a mechanisms by which a neonate's immune system is primed by the acquisition of immunoglobulins and antigens (Heyman and Desjeux, 1992).

When viewed under TEM, many of the sectioned particles showed an annular-like pattern of concentric rings. These rings and their positive staining with Alizarine red S, which highlights calcified tissue, suggests that the particles are bone in origin. The source of this bone is undoubtedly the skeleton of the rodent meals which the pythons have ingested intact. The rodent's skeleton is thoroughly broken down within the python's stomach (we have never observed bone fragments in the python's small intestine) and is passed into the small intestine with the rest of the meal (Secor, 2003). We expect that within the enterocytes the particles are degraded by lysosomes and what remains (calcium, phosphate, minerals, etc.) then passes into circulation.

An alternative explanation for the particles is that they are the cysts of parasites which were originally present in the rodent meals, and having escaped gastric digestion became embedded within the intestinal mucosa. We feel that this is unlikely considering that the rodents fed to our snakes were raised in institutional animal care facilities and thus should be parasite-free, that the cysts probably would not have survived the digestive actions of the python stomach, and that we would not expect the cysts to be stained by Alizarine red S.

*Further studies of the morphological response in pythons*

While the postprandial increases in mucosal mass, enterocyte width, and microvillus length have been well described, there is still much to be learned from studies of the python's intestine. Hence, we propose the following six questions to spark further exploration into the trophic responses of the python intestine to feeding and fasting.

*1) What is the mechanism of enterocyte hypertrophy?* Several mechanisms, including nutrient accumulation and fluid pumping, have been proposed to underlie the postprandial increase in enterocyte width. Determining which of these mechanisms (or both) is (are) responsible may require tracking labelled luminal or endogenous components (amino acids) to the enterocytes and observing their incorporation into the cells. Once identified, the mechanisms responsible for enterocyte hypertrophy for the python would be expected to also underlie the postprandial hypertrophy of enterocytes for other organisms.

*2) Does cellular hyperplasia also contribute to postprandial increases in intestinal mucosa?* The fact that mammals respond to increases in digestive demand by increasing rates of enterocyte proliferation (hyperplasia) (Aldewachi et al., 1975; Goodlad et al., 1988) suggest that pythons also do so. For the Burmese python, rates of enterocyte proliferation have been found to either increase (Secor et al., 2000) or remain stable (Starck and Beese, 2001) immediately after feeding. If rates of enterocyte apoptosis match rates of proliferation, cellular hyperplasia would contribute modestly to the increase in mucosal mass (Lignot and Secor, 2003). Studies that quantify the relative occurrence of proliferation and apoptotic events, as well as determine the life span of enterocytes during fasting and digestion, would provide insight in the contribution of hyperplasia to intestinal response, and the cyclic changes in cellular turnover with each meal.

*3) What is the mechanism of transepithelial lipid absorption?* What are the mechanisms by which those lipids enter blood circulation for the python. Either reformed triglycerides are packaged into chylomicrons which are released into lacteals of the villus core and enter the lymphatic system as is the case for mammals, or the triglycerides are absorbed intact and move directly into villus circulation thereby bypassing lymphatic passage. Although there is marginal evidence to support the latter mechanism, including the lack of chylomicrons in circulation and absence of lipids in lymphatics when levels of plasma triglycerides are high (S. Secor and D. Puppione, unpublished observations), this is a process in need of further investigation.

*4) What are the cellular mechanisms underlying the lengthening and shortening of the intestinal microvilli?* Discovering the mechanisms by which microvilli lengthen and then retract for the python will underscore the explanation on how they can regulate intestinal performance with feeding and fasting. Techniques that can pinpoint the position of microvillar and membrane proteins and quantify the relative expression of those proteins will be instrumental in tracking the movement and synthesis of those proteins during the lengthening and shortening process.

*5) What is the source and fate of the spherical particles embedded in the brush border membrane?* Whereas the evidence suggests that the spherical particles are bone in origin, it is not clear whether enterocytes actively engulf particles in a phagocytotic fashion or if luminal pressure drives the particles into the cells. If they are bone particles which become assimilated, then pythons are absorbing more calcium and minerals from their diets than most other terrestrial carnivores which do not digest bone. Therefore, do pythons sequester more calcium from their meals into their skeletons and/or experience higher rates of calcium turnover and excretion than other carnivores?

*6) To what extent are other gastrointestinal tissues of the python altered trophically with feeding and fasting?* In considering that digestion is a very integrative response, involving all aspects of the gastrointestinal tract, as well as associated organs (liver, pancreas, and kidneys), the impressive postprandial responses of the intestine would expectedly transcend to these other tissues.

**Acknowledgment**

We thank C. Arbiol, C. Bobino, C. Chevallier, R. Clemens, E. Roth, S. Sampogna, and B. Sjostrand for technical assistance in this study. We thank J. Diamond for financial support (National Institute of General Medical Sciences Grant GM-14772) during a portion of this study. Support for this study was also provided by the National Institute of Diabetes and Digestive and Kidney Diseases Research Service Award (DK-08878), Howard Hughes Medical Institute, CNRS Life Science Department, and the University Louis Pasteur.

Figure 1. Wet mass of the small intestine and of each intestinal third (proximal, middle, and distal) as a function of days after feeding for *Python molurus* following their consumption of rodent meals equalling 25% of snake body mass. For the python, feeding generates a doubling of intestinal mass. In this and following figures, error bars represent ±1 S.E.M. and are omitted if the S.E.M. is smaller than the width of the symbol for mean value.

Figure 2. Tips of intestinal villi of *Python molurus* viewed by conventional scanning electron microscopy. Images were taken of samples from fasted snakes (a, b, c), and at 1 day postfeeding (d, e), 3 days postfeeding (f, g, h), 6 days postfeeding (i), and at 14 days postfeeding (j). C: cracks; E: enterocytes; HE: hypotrophic enterocytes; MV: microvilli; V: intestinal villi. Bars: a, d, f, i, j: 50 µm; b, c: 5 µm; e, g, h: 10 µm.

Figure 3. Tips of intestinal villi of *Python molurus* viewed by environmental scanning electron microscopy. Images were taken of samples from fasted snakes (a, b, c), at 1 day postfeeding (d, e, f), 3 days postfeeding (g, h, i) and from anterior (a, d, g, j, k, l), middle (b, e, h) and distal segments (c, f, i) of the intestine. Note for fasted pythons the presence of mucosal bridges between villi, and droplets on the villus surface after feeding. Under continuous viewing, droplets fused (j, k, 1 minute between images), and if the samples are allowed to dehydrate the droplets remain (l). LD: lipid droplet, M: mucus bridge; V: intestinal villi. Bars: a, b, c: 100 µm; d, g, h, i, l: 200 µm; e, j, k: 50 µm.

Figure 4. Semithin sections of intestinal villi of fasting *Python molurus* (a) and at 1 day postfeeding (b). Note the tightly packed and pseudostratified orientation of enterocytes of the fasted python and the subsequent increase in enterocyte width resulting in a single non-

overlapping layer after feeding. BB: brushborder; C: chorion; L: lumen; LD: lipid droplet; ME: monostratified epithelium; N: nuclei; PE: pseudostratified epithelium. Bar: 40 µm.

Figure 5. Width (a), length (b), and volume (c) of intestinal enterocytes of *Python molurus* plotted as a function of days postfeeding for the proximal, middle, and distal thirds of the small intestine. Note the postprandial increase in enterocyte width and volume and decrease in length, and the decrease in enterocyte dimensions in the distal portion of the small intestine.

Figure 6. Width of the **mucosa and muscularis/serosa** layers of the proximal, middle, and distal small intestine plotted as a function of time postfeeding for *Python molurus*. Note the significant increase in mucosal width of the proximal and middle small intestine, and the lack of any feeding effects on distal mucosa and the serosal of the entire small intestine.

Figure 7. Lipid droplets concentrated within enterocytes of the proximal small intestine 1 day after feeding for *Python molurus* (a). By day 6 of digestion, droplets are reduced in size and present along the basolateral membrane (b). The distal intestine lacking lipid droplets housed a large population of goblet cells (c). C: chorion; L: lumen; LD: lipid droplet; MC: mucus cell; RBC: red blood cell. Bar: 40 µm.

Figure 8. Electron micrographs of proximal intestinal microvilli of *Python molurus* fasted (a) and at 0.25 (b), 0.5 (c), 1 (d), 3 (e), 6 (g), and 14 (g) days postfeeding. Note the rapid postprandial lengthening of the python's microvilli and subsequent shortening following the completion of digestion. Bar: 1 µm.

Figure 9. Length (a), width (b), density (c), and surface area (d) of the intestinal microvilli of *Python molurus* as a function of time postfeeding. Complete profiles are presented for the microvilli of the proximal intestine and measurements from the middle and distal regions are illustrated for fasted pythons and at 1 and 6 days postfeeding. Density is quantified as the number of microvilli extending from a $\mu m^2$ of apical surface. Surface area represents the summed surface area of microvilli (calculated based on a formula for a cylinder) for a $\mu m^2$ of apical surface. Note the 5-fold increase in microvillus length and 4-fold increase in surface area soon after feeding.

Figure 10. Spheroid particles embedded in the brush border of the proximal and middle regions of the small intestine of digesting *Python molurus* as detected by light microscopy (a, b), and TEM (c). The bright staining of the particles by Alizarine red S suggest strongly that they are composed of calcium (b). Bars: a: 40 µm; b: 200 µm; c: 5 µm.

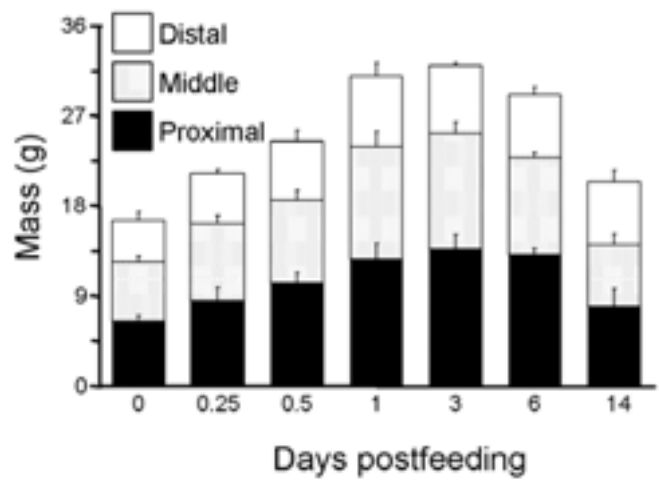

Days postfeeding

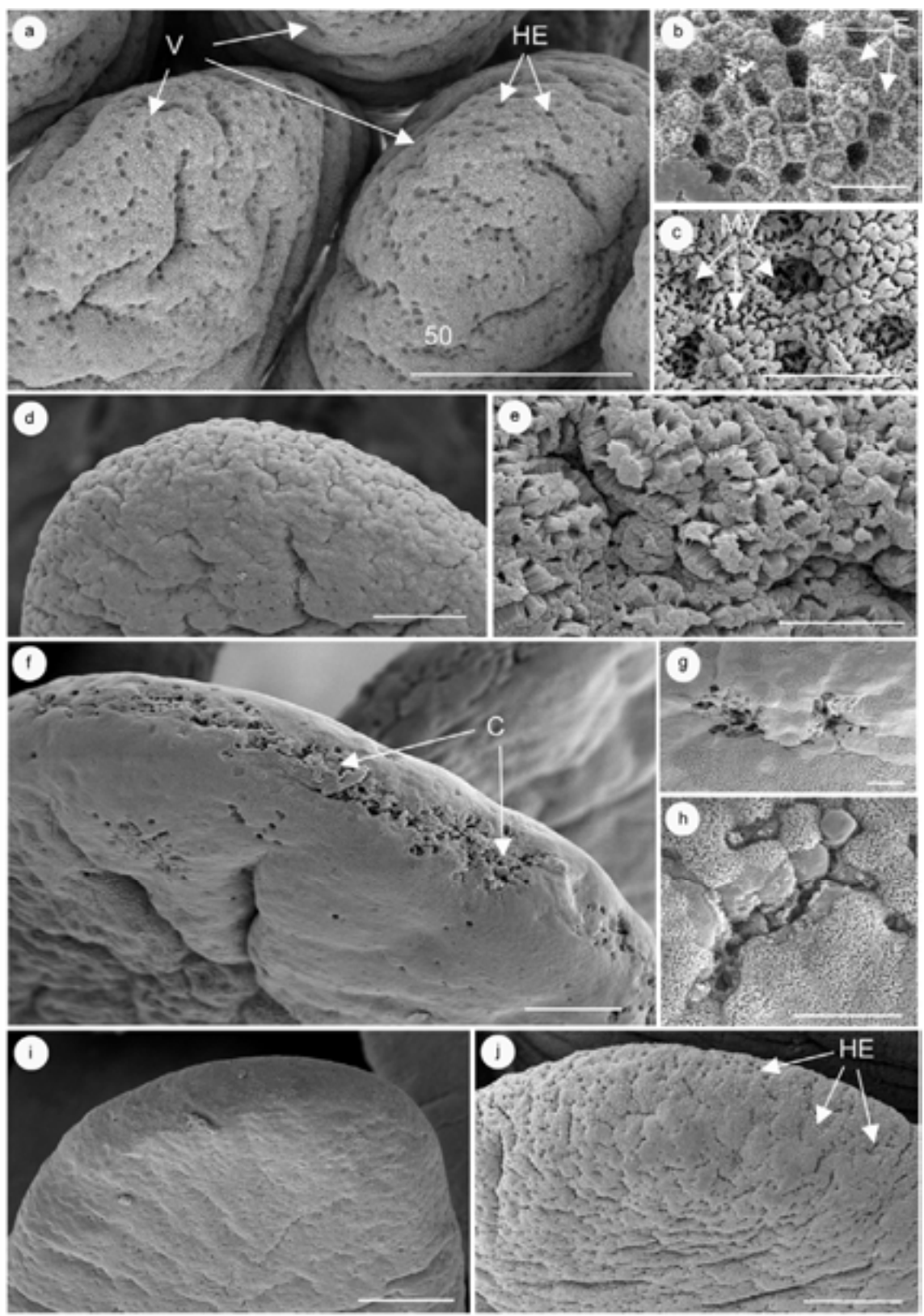

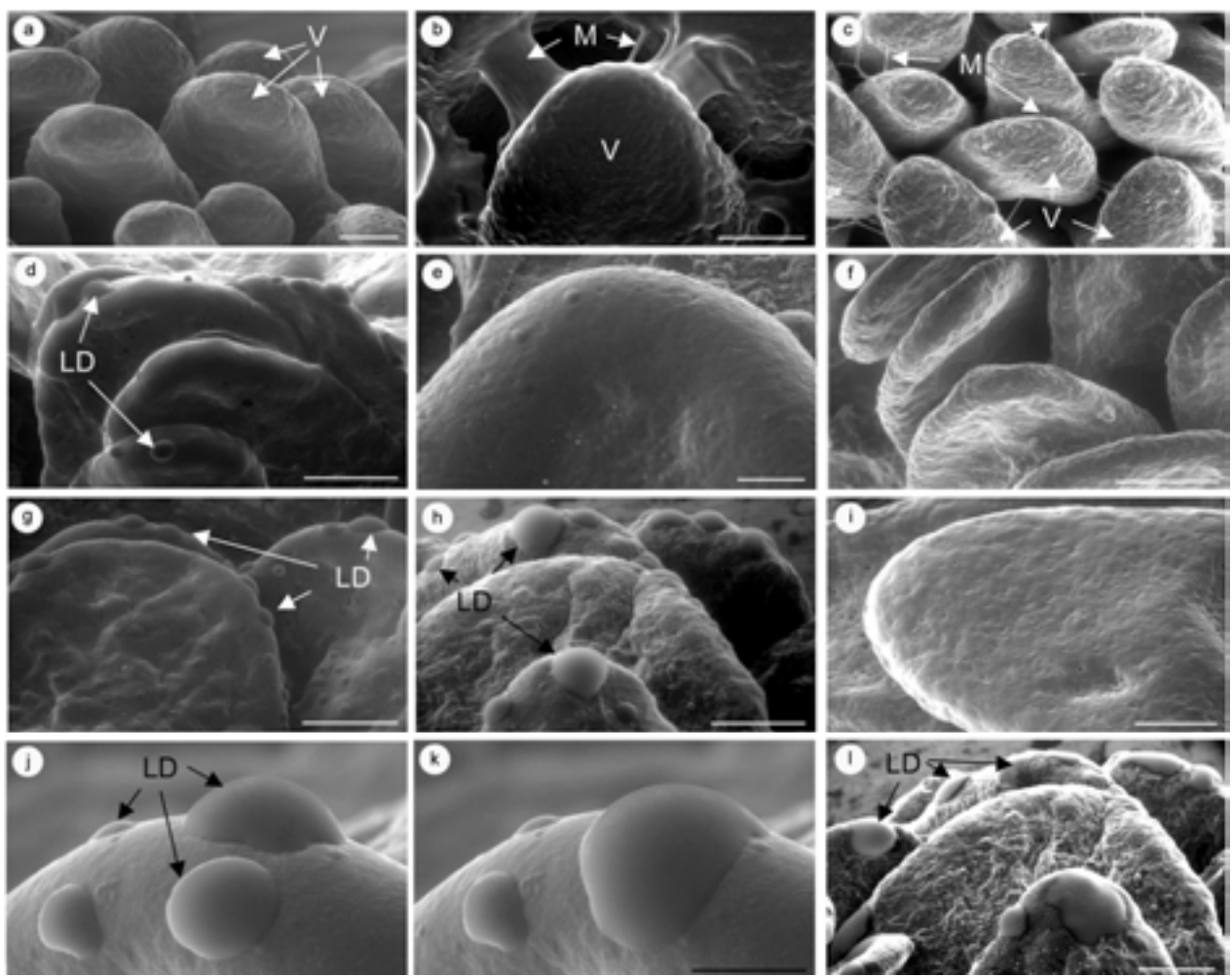

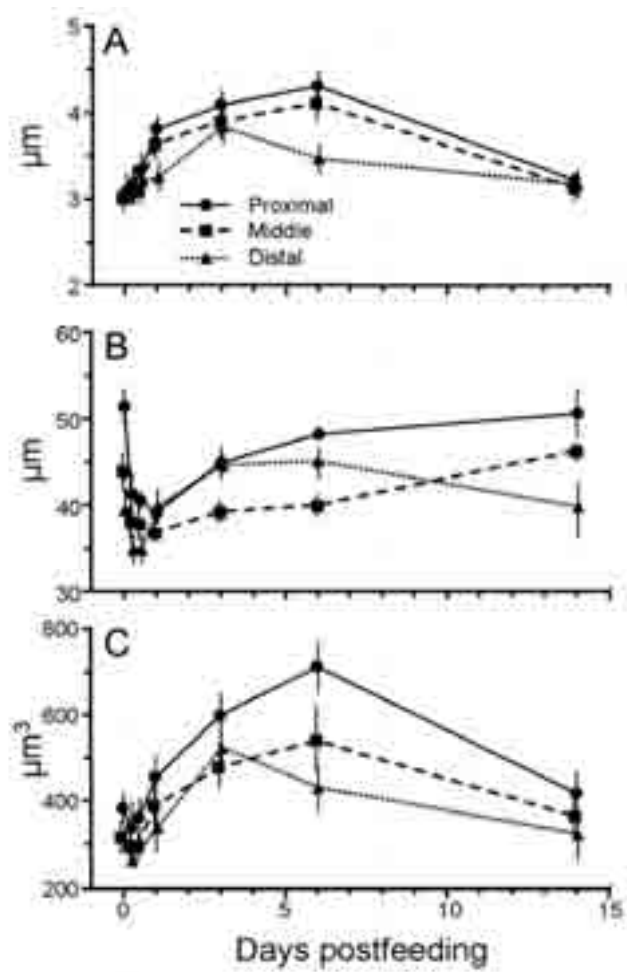

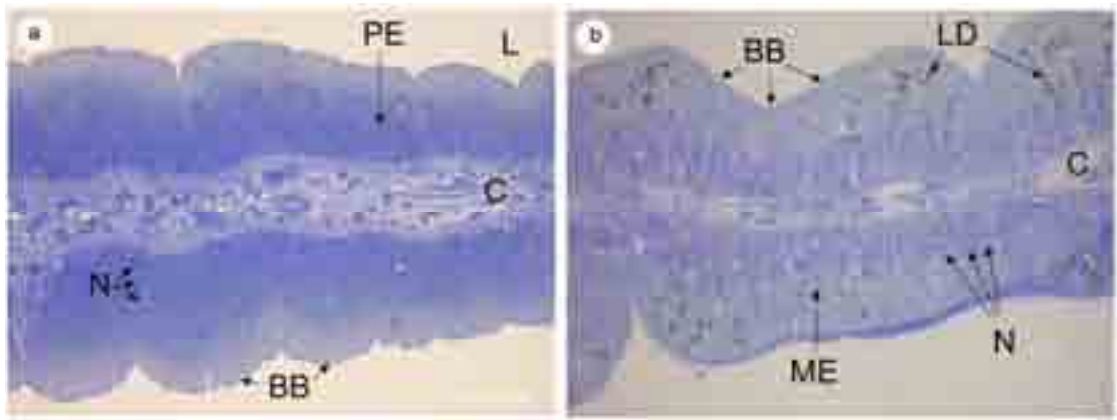

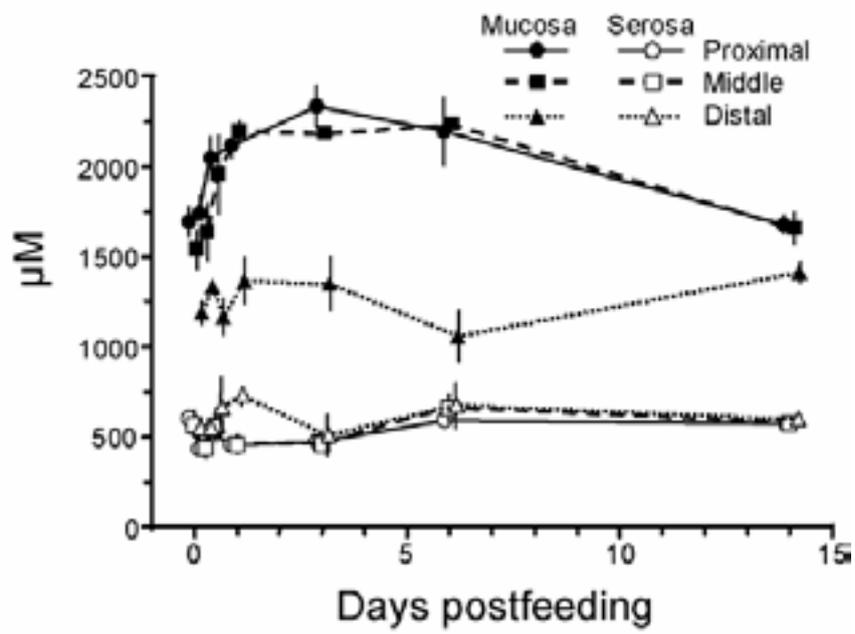

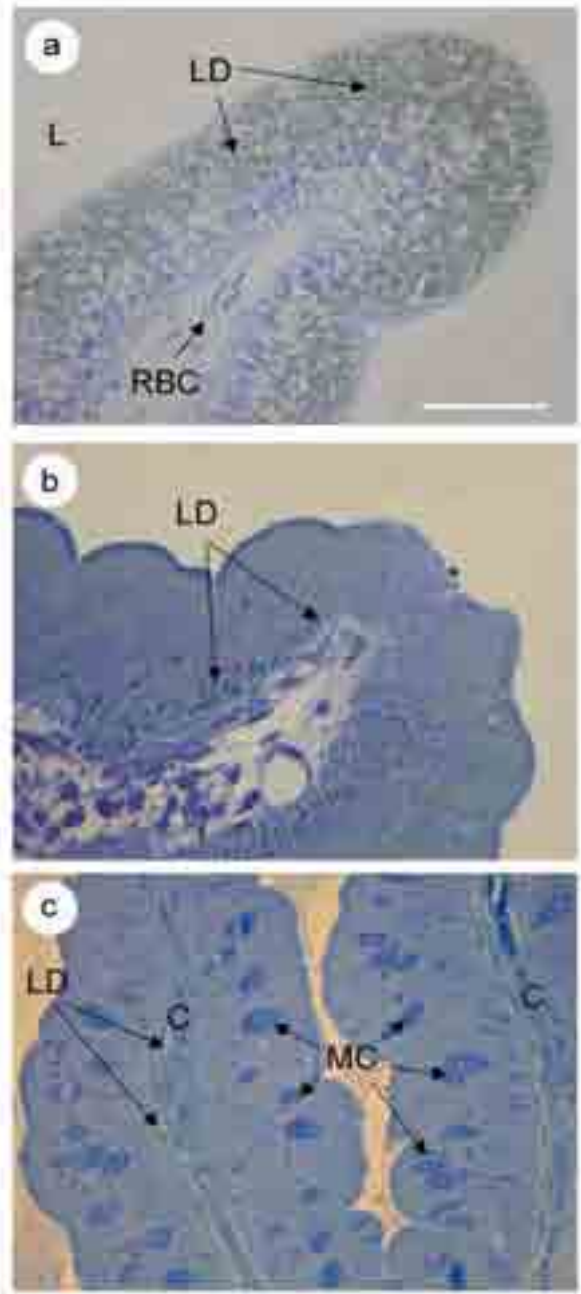

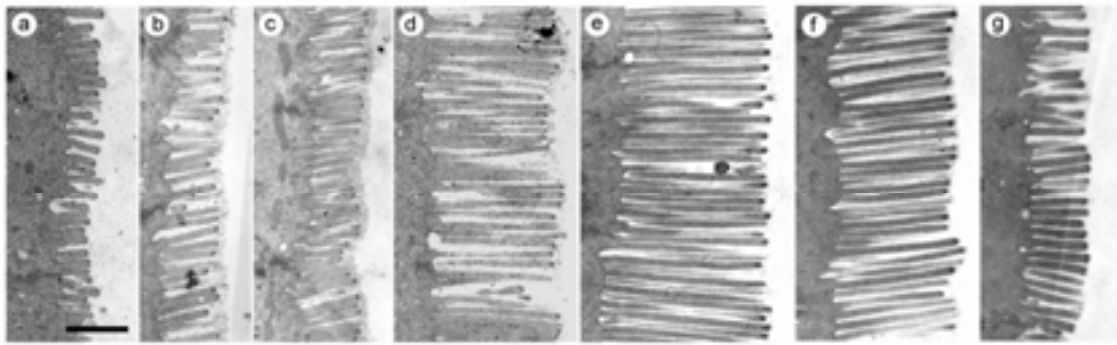

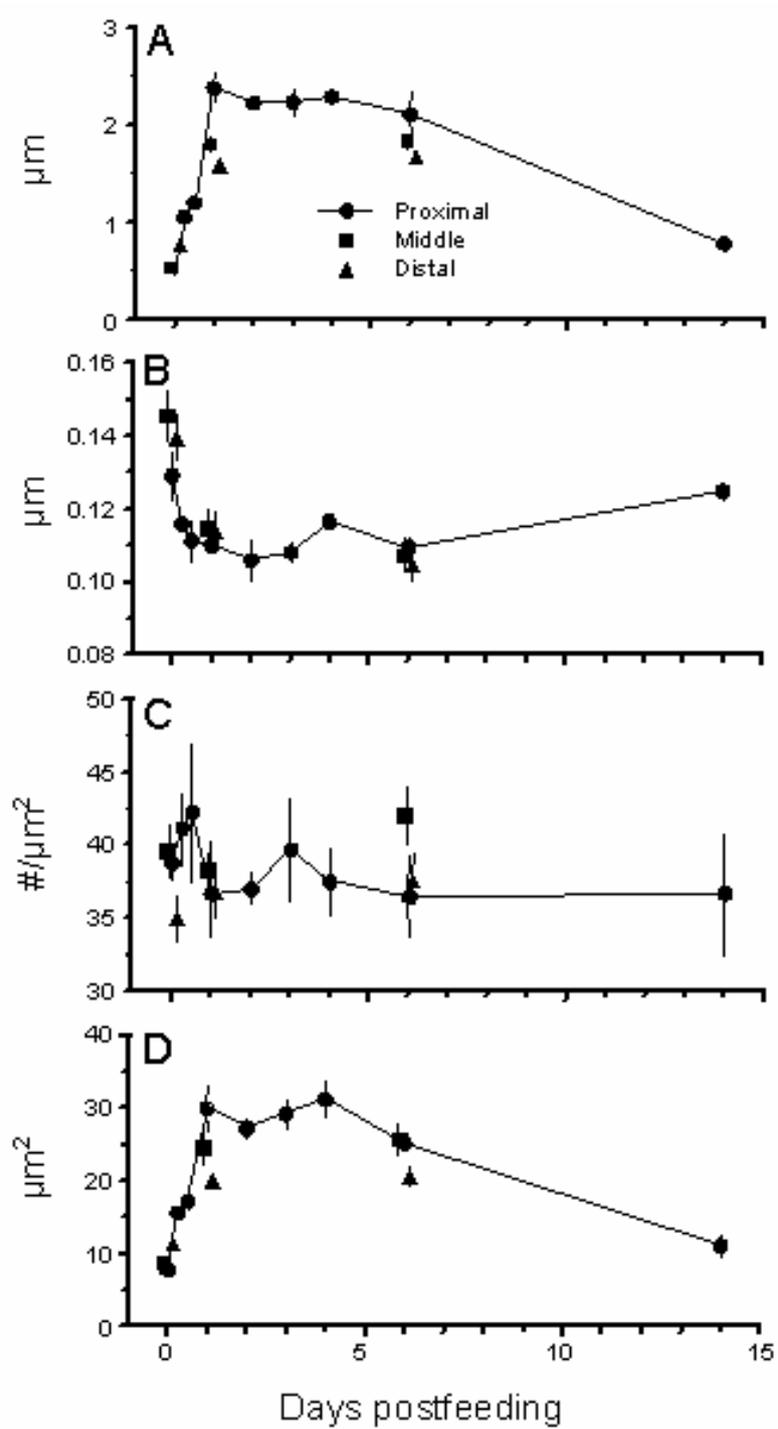

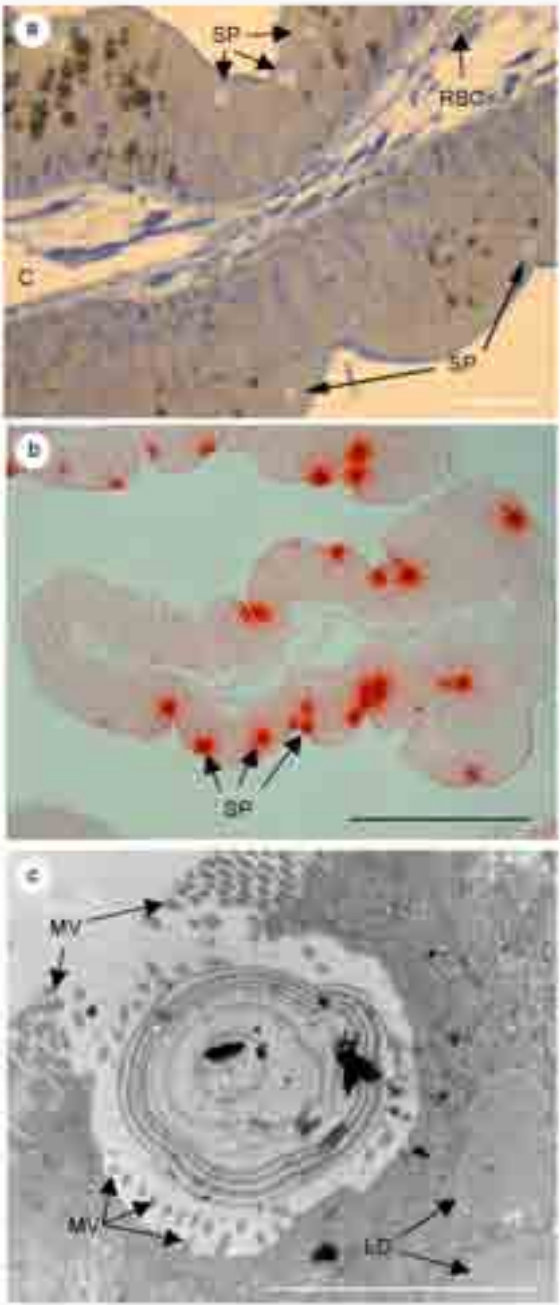